\documentclass[singlecolumn,epjc3]{svjour3}
\smartqed  
\RequirePackage{graphicx}
\journalname{Eur. Phys. J. C}
\begin{document}

\title{Possible existence of wormholes in the galactic halo region}

\author{Farook Rahaman\thanksref{e1,addr1}
\and
        P.K.F. Kuhfittig\thanksref{e2,addr2}
       \and
       Saibal Ray\thanksref{e3,addr3}
        \and
        Nasarul Islam\thanksref{e4,addr4}.
}

\thankstext{e1}{e-mail: rahaman@iucaa.ernet.in}
\thankstext{e2}{e-mail: kuhfitti@msoe.edu}
\thankstext{e3}{e-mail: saibal@iucaa.ernet.in}
\thankstext{e4}{e-mail: nasaiitk@gmail.com}

\institute{Department of Mathematics, Jadavpur University, Kolkata
700 032, West Bengal, India\label{addr1}
              \and
              Department of Mathematics, Milwaukee School of Engineering, Milwaukee, Wisconsin
53202-3109, USA\label{addr2}
              \and
              Department of Physics, Government College of Engineering \& Ceramic Technology, Kolkata
700 010, West Bengal, India\label{addr3}
              \and
               Department of Mathematics, Danga High Madrasah, Kolkata 700 103, West Bengal,
India\label{addr4} }

\date{Received: date / Accepted: date}

\maketitle

\begin{abstract}
 Two observational results, the density
profile from simulations performed in the $\Lambda$CDM scenario
and the observed flat galactic rotation curves, are taken as input
with the aim of showing that the galactic halo possesses some of
the characteristics needed to support traversable wormholes.  This
result should be sufficient to provide an incentive for scientists
to seek observational evidence for wormholes in the galactic halo
region.
\end{abstract}



\section{Introduction}

In recent years observational evidence has been found for black
holes, once considered to be hypothetical astrophysical objects.
An interesting challenge is to find evidence for another type of
strange object, the traversable wormhole, a tunnel-like structure
connecting different regions of our Universe or of different
universes altogether. Although just as good a prediction of
Einstein's theory as black holes, they have so far eluded
detection. Unlike black holes, holding a wormhole open requires
the violation of the null energy condition, an example of which is
the Casimir effect \cite{MT88}.  On the cosmological level,
phantom dark energy also violates the null energy condition and
could therefore give rise to wormholes \cite{fL05,pK09}.

Moving to the galactic level, we are confronted with other
peculiar phenomena that cannot be explained by the standard
model, examples of which are the observed flat rotation curves
in galaxies.  In particular, the rotation curves of neutral
hydrogen clouds in the outer regions cannot be explained in
terms of ordinary (luminous) matter.  These phenomena have
led to the hypothesis that galaxies and even clusters of
galaxies are pervaded by some non-luminous matter, now called
\emph{dark matter}.  Dark matter is able to account for these flat
rotation curves.  The term  \emph{dark} refers to the fact that
it does not emit electromagnetic waves, nor interact with
normal matter.  A number of candidates for dark matter have
been proposed over time: new particles predicted by
supersymmetry \cite{JKG96}, massive neutrinos collectively
known as WIMPs (weakly interacting massive particles)
\cite{KT90}, a source of scalar fields \cite{sF04}, global
monopoles \cite{NSS00}, brane-world effects of gravitation
\cite{MH04}, noncommutative geometry \cite{RKCUR12},
geometric effects of $f(R)$ gravity \cite{BHL08}, $f(T)$
gravity \cite{fR12}, etc.

To see how wormholes might fit in with these strange astrophysical
phenomena, we begin by noting that Navarro et al. \cite{jN96} have
used $N$-body simulations to search out the structure of dark
halos, in particular the density profile of dark halos in the
standard CDM cosmology.  Their numerical simulations in the
$\Lambda$CDM scenarios led to the density profile of galaxies and
clusters of galaxies having the form
\begin{equation}\label{E:rho}
\rho(r)=\frac{\rho_s}{\frac{r}{r_s}\left( 1+\frac{r}{r_s}
\right)^2},
\end{equation}
where $r_s$ is the characteristic scale radius and $\rho_s$ the
corresponding density.  Since this density profile of CDM halos of
several masses (between $3\times 10^{11} M_\odot$ and $3\times
10^{15} M_\odot$) fits accurately, we will rely on Eq. (\ref{E:rho})
to show that the galactic halo may be able to support traversable
wormholes.

In the present work, essentially we are motivated to show that the
geometry of the space-time of a galactic halo may be described by
a traversable wormhole metric, fitting with the expected density
profile predicted by simulations and with the observed flat
galactic rotation curves.

\section{The solutions}

While we now have the density profile, other properties of
dark matter remain unknown.  We will therefore assume that
dark matter has the most general anisotropic energy-momentum
tensor given by
\begin{equation}
 T_\nu^\mu=(\rho + p_t)u^{\mu}u_{\nu} - p_t g^{\mu}_{\nu}+
            (p_r -p_t )\eta^{\mu}\eta_{\nu},
\end{equation}
with $u^{\mu}u_{\mu} = - \eta^{\mu}\eta_{\mu} = 1$, $p_t$ and $p_r$
being the transverse and radial pressures, respectively.

As noted earlier, the observed flat rotation curves of neutral
hydrogen clouds in the outer regions of galaxies indicate the
existence of dark matter. In such galaxies these neutral hydrogen
clouds are therefore treated as test particles moving in circular
orbits. The spacetime in the galactic halo is characterized by the
line element
\begin{equation}
ds^2=-e^{2f(r)}dt^2+e^{2g(r)}dr^2
+r^2(d\theta^2+\sin^2\theta\,d\phi^2).
\end{equation}
A more convenient form for later analysis is
\begin{equation}\label{E:line1}
ds^2=-e^{2f(r)}dt^2+\left(1-\frac{b(r)}{r}\right)^{-1}dr^2
+r^2(d\theta^2+\sin^2\theta\,d\phi^2).
\end{equation}
A flat rotation curve for the circular stable geodesic motion
in the equatorial plane yields
\begin{equation}\label{E:redshift}
 e^{2f(r)}= B r^l,
\end{equation}
derived in Appendix 1. Here $l = 2(v^\phi)^2$, where $v^{\phi}$ is
the rotational velocity and $B$ is an integration constant. The
observed rotation curve profile in the dark matter region
indicates that the rotational velocity $v^\phi$  is nearly
constant. For example, for a typical galaxy of mass $1.8 \times
10^{12} M_\odot$ within 300 kpc \cite{NSS01}, the rotational
velocity is $v^\phi \sim 10^{-3}$~(300~km/s). So by letting
$B=1/r_s^l$, the spacetime metric becomes
\begin{equation}\label{E:line2}
  ds^2=-\left(\frac{r}{r_s}\right)^ldt^2
     +\left(1-\frac{b(r)}{r}\right)^{-1}dr^2+r^2(d\theta^2
     +\sin^2\theta\,d\phi^2).
\end{equation}

As shown in Appendix 2, the Einstein field equations
$(G_{\mu\nu}=8\pi T_{\mu\nu})$ now yield
\begin{equation}
b(r)= 8 \pi \rho_s r_s^3 \left[\ln \left( 1+\frac{r}{r_s} \right) +
\frac{1}{\left( 1+\frac{r}{r_s} \right)} \right]
\end{equation}
and
\begin{eqnarray}
8 \pi   r_s^2  p_r= \frac{ l}{(\frac{r}{r_s})^2} \left[ 1- \frac{ 8
\pi \rho_s r_s^2}{\frac{r}{r_s}} \left\{\ln \left( 1+\frac{r}{r_s}
\right) + \frac{1}{\left( 1+\frac{r}{r_s} \right)} \right\} \right]
\nonumber
\\-\frac{ 8 \pi \rho_s r_s^2}{(\frac{r}{r_s})^3}
\left[\ln \left( 1+\frac{r}{r_s} \right) + \frac{1}{\left(
1+\frac{r}{r_s} \right)} \right].
\end{eqnarray}
It should be emphasized that this result is based on the two
cosmological observations made earlier, the density profile, Eq.
(\ref{E:rho}), and the observed rotation curve profile.
(The expression for the transverse pressures is given in
Appendix 2.)

Having obtained both $f(r)$ and $b(r)$, we are now in a position
to examine the spacetime metric more closely.  First recall that
if the line element, Eq. (\ref{E:line1}), is to represent a
wormhole, then\\
1) The \emph{redshift function}, $f(r)$, must remain finite to
prevent an event horizon. \\
2) The \emph{shape function}, $b(r)$, must obey the following
conditions at the throat $r = r_0$ :  $b(r_0) = r_0$ and
$b^\prime(r_0) < 1$, the so-called flare-out condition.\\
3) $b(r)/r < 1$ for $r >r_0$.\\

\begin{figure*}
\begin{tabular}{rl}
\includegraphics[width=5.5cm]{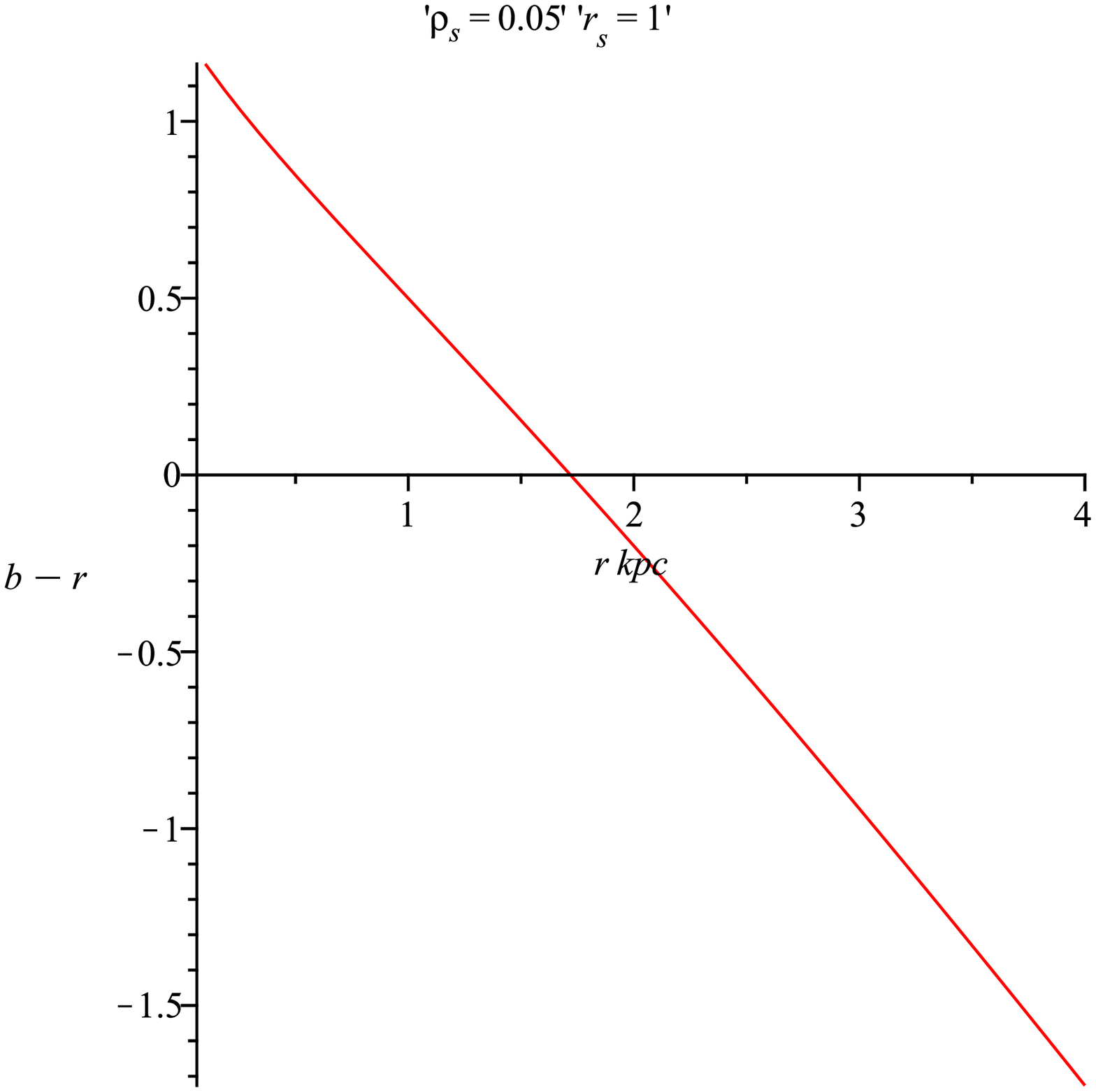}
\includegraphics[width=5.5cm]{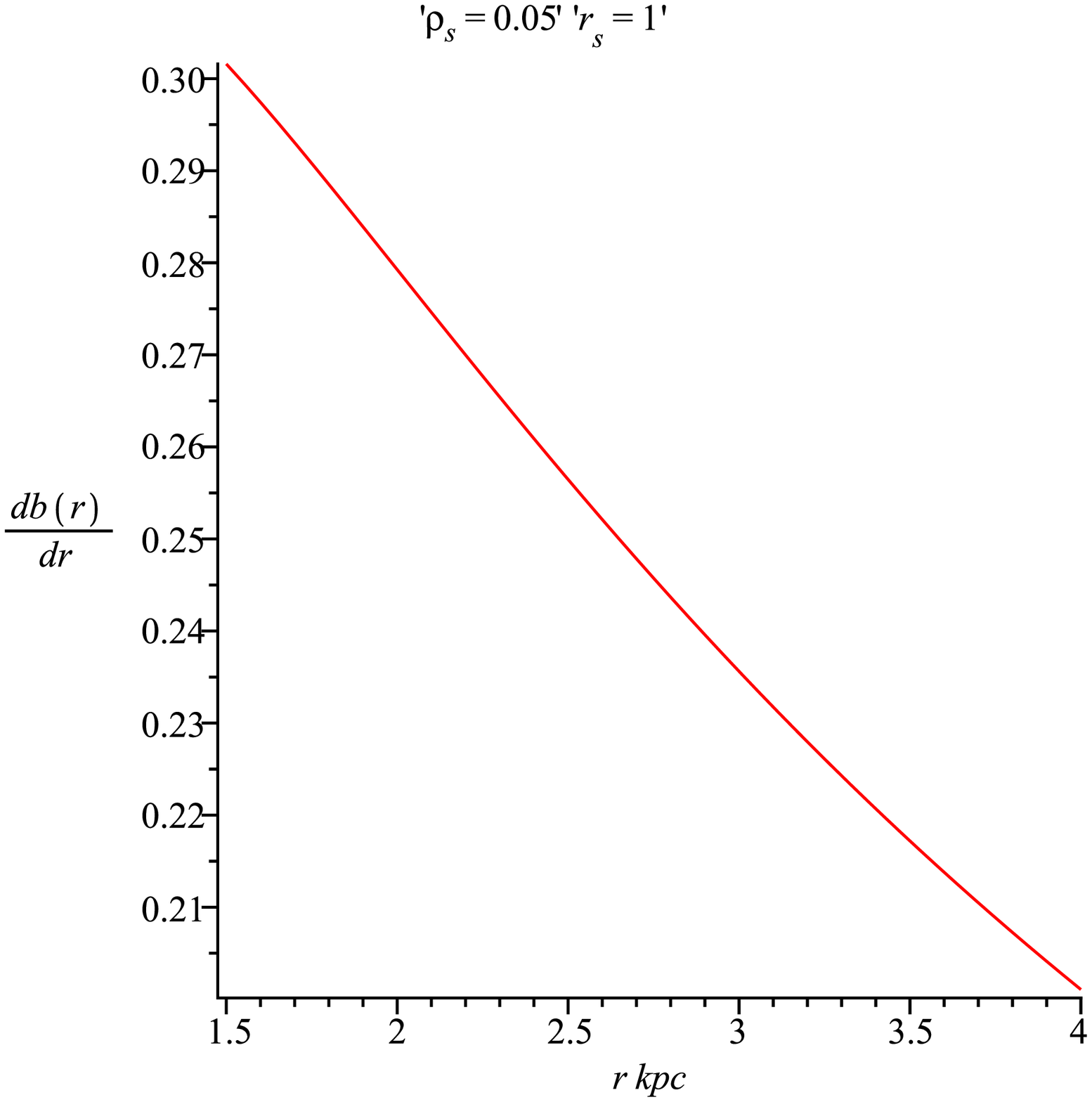}&
\includegraphics[width=5.5cm]{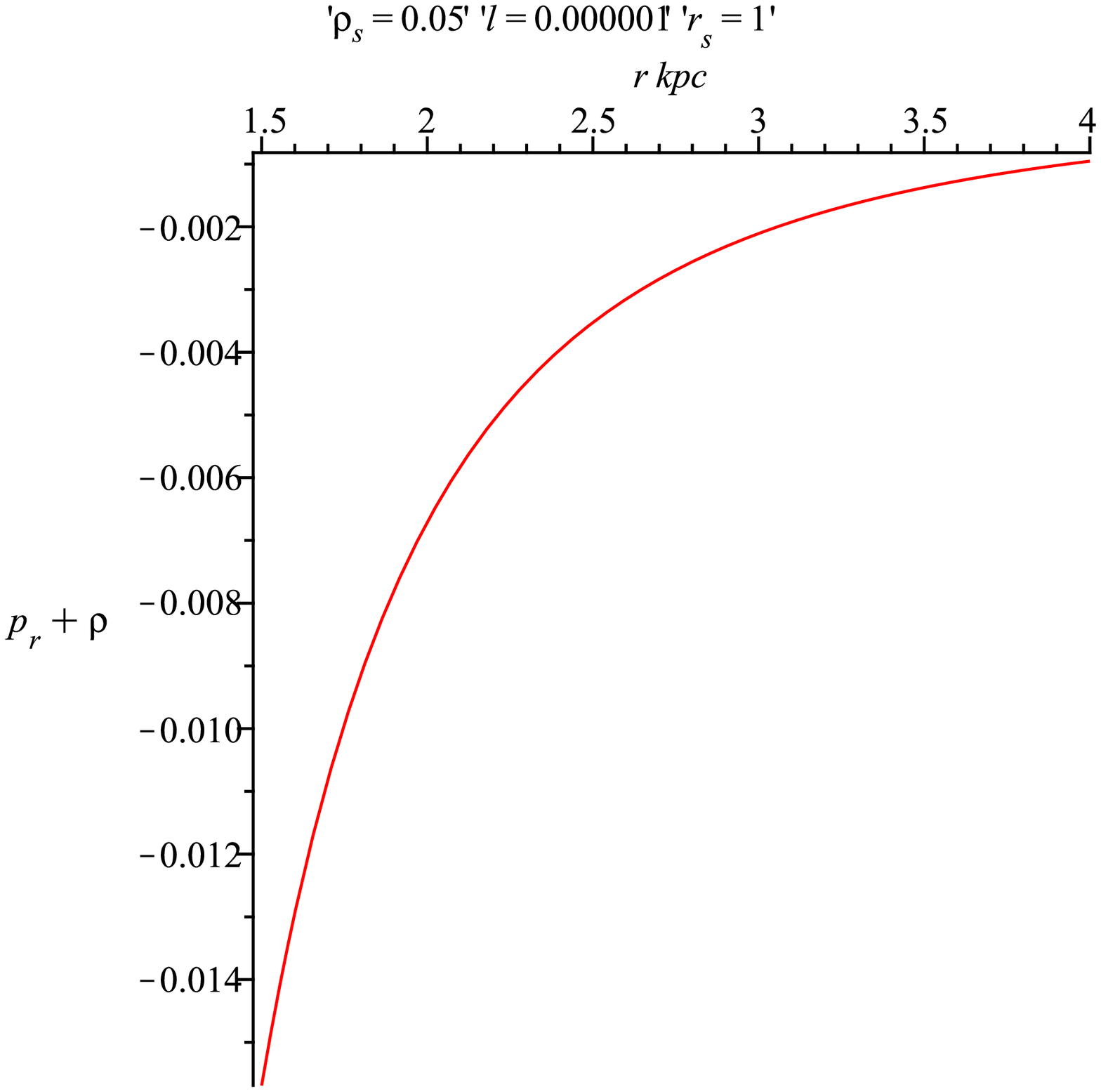} \\
\end{tabular}
\caption{ (\textit{Left})  The throat of the wormhole occurs where
$b(r)-r$ cuts the $r$-axis. (\textit{Middle})  Diagram of the
derivative of the shape function of the wormhole. (\textit{Right})
The variation of the left-hand side of the expression for the null
energy condition of matter in the galactic halo is plotted against
$r$. We have used geometric units, G=c=1 for choosing the values
of $\rho_s = 0.05$ and $r_s =1$.}
\end{figure*}

Regarding these requirements, observe that Eq. (\ref{E:line2})
shows that the spacetime does not have an event horizon.  To check
the shape function, we will use a graphical approach by using some
typical values of the parameters.  Fig. 1 (left panel) shows the
following: the throat is located at $r=r_0$, where $b(r)-r$ cuts
the $r$-axis.  Also,  for $r>r_0$, $b(r)-r<0$, which implies that
$b(r)/r<1$, an essential requirement for a shape function.
Moreover, $b(r)-r$ is a decreasing function for $r\ge r_0$.
Therefore, $b'(r_0)<1$, so that the the flare-out condition is
satisfied.  Fig. 1 (middle panel) also supports this assertion. So
all three conditions are satisfied.  For the sake of completeness,
observe that for the values in Fig. 1, $\rho_s =0.05$ and $r_s =1
$, we obtain $r_0=1.7192$ kpc  to four decimal places with
$b'(1.7192)\approx 0.29218$.

Our final task concerning the wormhole structure is to examine
the null energy condition.  This condition must be violated if
the wormhole is to remain open \cite{MT88}.  Judging from
Fig. 1 (right panel), this is indeed the case since
$p_r+\rho<0$.

For a spacetime to be asymptotically flat, both $f(r)$ and
$b(r)/r$ have to approach zero as $r\rightarrow \infty$.  The
second condition is satisfied, but not the first, as we can
see from Eq. (\ref{E:line2}).  So the wormhole cannot be
arbitrarily large, which also applies to the halo region.
The usual procedure is to cut off the wormhole material
at some radial distance and join the solution to an external
Schwarzschild spacetime.

It is also useful to calculate the active gravitational mass
of the wormhole from the throat, $r_0$ (in kpc) up to the
radius $R$.  This mass is given by
\begin{equation}
\label{eq40} M_{active}=4\pi\int_{r_0}^{R} \rho r^2 dr\\
= 4 \pi \rho_s r_s^3  \left [\ln \left( 1+\frac{r}{r_s} \right) +
\frac{1}{\left( 1+\frac{r}{r_s} \right)} \right]^R_{r_0}.
\end{equation}
Observe that the active gravitational mass $M_{active}$
of the wormhole is positive.  This implies that seen from
the Earth, we would not be able to distinguish the gravitational
nature of a wormhole from that of a compact mass in the galaxy.

\section{Equilibrium condition}

The generalized Tolman-Oppenheimer-Volkov (TOV) equation is
\begin{equation}
  \frac{dp_r}{dr}   +
\frac{\nu^\prime}{2}\left(\rho +p_r\right) +
\frac{2}{r}\left(p_r - p_t\right) = 0.
\end{equation}

According to Ponce de Le\'{o}n's suggestion \cite{Leon1993}, we rewrite the above  TOV equation (10)
 for the anisotropic  mass distribution in the galactic halo, to the following form
\begin{equation}
-\frac{M_G\left(\rho +p_r \right)}{r^2}e^{\frac{\lambda-\nu}{2}}-\frac{dp_r }{dr}
+\frac{2}{r}\left(p_t -p_r \right)=0,
\end{equation}
where $M_G=M_G(r)$ is the effective gravitational mass from the throat to some
radius $r$ and is given by
\begin{equation}
M_G(r)=\frac{1}{2}r^2e^{\frac{\nu-\lambda}{2}}\nu^{\prime}.
\end{equation}
This expression of mass   can be derived from the Tolman-Whittaker formula and
the Einstein's field equations. It is quite natural that  the modified TOV
equation (11) provides the information of   the equilibrium condition for the wormhole
subject to gravitational ($F_g$) and hydrostatic ($F_h$) plus
another force due to the anisotropic nature ($F_a$) of the matter
comprising the wormhole.  Hence,  for equilibrium the above equation (11)  takes the form
\begin{equation}
 F_g+ F_h + F_a=0,\label{eq33}
\end{equation}
where,
\begin{equation}
F_g =-\frac{\nu^\prime}{2}\left(\rho +p_r\right)
\end{equation}
\begin{equation}
F_h =-\frac{dp_r}{dr}
\end{equation}
\begin{equation}
F_a=\frac{2}{r}\left(p_t -p_r\right)
\end{equation}
The profiles of $F_g$, $F_h$ and $F_a$ for the matter distribution of the galactic halo region are
shown in Fig. 2. The figure indicates that equilibrium stage can
be achieved due to the combined effect of pressure anisotropic,
gravitational and hydrostatic forces. It is to be   noted
that value of $F_g$ is too small.  The other two
plots reside nearly opposite to each other to make the system balanced.

\begin{figure}[htbp]
\centering
\includegraphics[scale=.3]{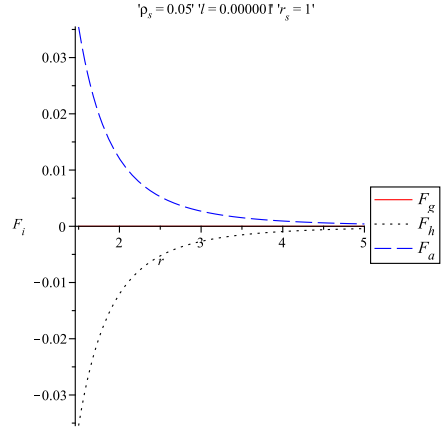}
\caption{Three different forces acting on fluid elements in static
equilibrium is shown against $r$. Value of $F_g$ is too small. \label{fig5}}
\end{figure}

\section{Scattering of scalar waves in wormhole geometry}

The minimally coupled massless wave equation in a wormhole
background is given by
\begin{equation}
 \qed \Phi = \frac{1}{\sqrt{-g}} \partial_\mu[\sqrt{-g}g^{\mu \nu}\partial_\nu \Phi ] =0.
\end{equation}

Note that for simplicity, we are dealing with minimally coupled
scalar waves. Since the wormhole spacetime is spherically
symmetric, the equation related to scalar field can be solved by
separation of variables,
\begin{equation}
  \Phi_{l_0m}  = Y_{l_0m}(\theta,\phi)\frac{U_{l_0}(r,t)}{r}.
\end{equation}
Here $Y_{l_0m}(\theta,\phi)$ are the spherical harmonics and
$l_0$ is the quantum angular momentum.

The possibility of astrophysical observations now provides the
motivation for studying the scattering of scalar waves in our
wormhole spacetime. Such observations would be important for
research on the gravitational radiation, as well as for
determining the possible existence of actual physical wormholes.

Using the separable form (18) in (17), one can obtain
\begin{equation}
 \left[ \frac{1}{\sin \theta}  \frac{\partial}{\partial \theta} \sin \theta \frac{\partial}{\partial
 \theta}+
 \frac{1}{\sin^2 \theta}  \frac{\partial^2}{\partial \phi^2}\right]
 Y_{l_0m}= l_0(l_0+1) Y_{l_0m}
\end{equation}

and

\begin{equation}
 \ddot{U}_{l_0} + \frac{\partial^2U_{l_0}}{\partial {r^*}^2}  = V_{l_0} U_{l_0},
\end{equation}
where the potential $V_{l_0}$ is given by
\begin{equation}
 V_{l_0} = e^{2f} \left[  \frac{l_0(l_0+1)}{r^2} -
 \frac{b'r-b}{2r^3}+\frac{1}{r}\left(1-\frac{b}{r}\right)f'\right].
\end{equation}
Here we have used the tortoise coordinate transformation $r^*$,
i.e.,
\begin{equation}
  \frac{\partial }{\partial {r^*}}  = e^f \sqrt{ 1-\frac{b}{r} }\frac{\partial }{\partial
  r},
\end{equation}
where the dot represents the differentiation with respect to $t$.
Actually, $r^*$ is the proper distance given by (using $r_s =1$)
\begin{equation}
  r^*  = \int_{r_0}^r \frac{x^{-\frac{l}{2}} dx}{\sqrt{1-\frac{4 \pi \rho_s \ln (1+x) + \frac{4 \pi \rho_s}{1+x}}{x}}} .
\end{equation}
Since integration cannot be performed in exact analytical form, we
find the numerical values of the proper distance $r^*$ for given
values of radial distance $r$ from the throat radius $r_0$, which
is shown in Table 1.

\begin{table}
\begin{center}
\caption{Values of $r^*$  for different r. ($r_0 = 1.7192$,
$r_s=1$ , $\rho_s = .05$) }\label{tab:1} {\begin{tabular}{ll}
\hline\noalign{\smallskip} $r$ & $r^*$
\\
\noalign{\smallskip}\hline\noalign{\smallskip} 5 & 4.9884 \\ 10 &
11.1906  \\ 15 & 16.9754  \\ 20 & 22.5755  \\ 25 & 28.0675  \\ 30
& 33.4974  \\ \noalign{\smallskip}\hline
\end{tabular}}
\end{center}
\vspace*{5cm}
\end{table}

Observe that characteristics of the potential are determined by
the shape and redshift functions of the wormhole.

Assuming the time dependence of the wave to be harmonic, one can
write
\begin{equation}
 U_{l_0} (r,t)  = \widehat{U}_{l_0} (r,\omega)e^{-i\omega t}.
\end{equation}
Using (17) in (20),  we get the Schr\"{o}dinger equation
\begin{equation}
\left[\frac{d^2}{d{r^*}^2} +\omega^2-  V_{l_0}(r) \right]
\widehat{U}_{l_0} (r,\omega) = 0.
\end{equation}

Near the throat ($r_0\longrightarrow b(r_0)$), the potential
$\approx e^{2f(r_0)} \left[  \frac{l_0(l_0+1)}{r_0^2}\right]$,
which is finite.

Since the wormhole proposed here is not arbitrarily large, we
assume that the wormhole material extends from the throat
$r_0=1.7192$ kpc to the radius $300$ kpc. For the value of $l=
.000001$, note that the magnitude of $V_{l_0}$ is negligible at
$r=30$ kpc. This means that the solution has the form of a plane
wave $ \widehat{U}_{l_0}\sim e^{\pm i \omega r^*}$ at the
distance $r=300$ kpc. This result indicates that if a scalar wave
passes through the wormhole, the solution would be changed from
$e^{\pm i \omega r}$ to $e^{\pm i \omega r^*}$. This confirms
that the potential affects the scattering of scaler waves.

\section{Conclusion}
We have shown in this paper that the galactic halo possesses some
of the characteristics needed to support a traversable wormhole.
The analysis is based on two observational results, the density
profile from simulations performed in the $\Lambda$CDM scenario
and the observed flat galactic rotation curves.  The results
should provide sufficient incentives for scientists to seek
observational evidence for wormholes, all the more since our study
is based on the rotational velocity $v^\phi \sim
10^{-3}$~(300~km/s) and a mass of $1.8 \times 10^{12} M_\odot $
within 300  kpc, making our own galaxy typical enough to be a good
candidate. We have briefly studied here balancing of the forces
that provides the equilibrium configuration of the system and also
proposed a possible detection of such wormholes by studying the
scattering of scaler waves.

\vspace{1.0cm}

\noindent \textbf{Appendix 1}

We derive the tangential velocity of circular orbits for the line
element
\begin{equation} ds^{2}=-e^{\nu(r)}dt^{2}+e^{\lambda(r)}dr^{2}
+r^{2}(d\theta^{2} + sin^{2}%
\theta \,d\phi^{2}). \end{equation}
The Lagrangian for a test particle is given by
\begin{equation}  2 \mathcal{L}  = -e^{\nu(r)}\dot{t}^{2}
+e^{\lambda(r)}
\dot{r}^{2}+r^{2}(\dot{\theta}^{2}+ sin^{2}%
\theta \,\dot{\phi}^{2}), \end{equation} where the overdot
indicates differentiation with respect to the affine parameter
$s$. The metric coefficients do not depend explicitly on $t$,
$\theta$, or $\phi$.  So the Euler-Lagrange equation yields
directly the following conserved quantities: the energy $E =-
e^{\nu(r)}\dot{t}$, the $\theta$-momentum $L_{\theta}=
r^2\dot{\theta}$, and the $\phi$-momentum  $L_\phi = r^2 {sin}^{2}
\theta\, \dot{\phi}$.  So the square of the total angular momentum
is
\begin{equation}\label{E:Lsquared}
   L^2=L_{\theta}^2+\left(\frac{L_{\phi}}{sin{\theta}}
   \right)^2=r^4(\dot{\theta}^2+ sin^2\theta\,\dot{\phi}^2).
\end{equation}

With the conserved quantities $E$ and $L$ and the norm of the
four-velocity $u^{\mu}u_{\nu}=-1$, the geodesic equation becomes
\begin{equation}\label{E:geodesic}
   -1=-e^{\nu(r)}\dot{t}^2+e^{\lambda(r)}\dot{r}^2+
   r^2(\dot{\theta}^2+{sin}^2\theta\,\dot{\phi}^2).
\end{equation}
As a result,
\begin{equation}
  e^{\nu(r)+\lambda(r)}\dot{r}^2+e^{\nu(r)}\left(1+
  \frac{L^2}{r^2}\right)=E^2
\end{equation}
or
\begin{equation}\label{E:eff1}
   e^{\lambda(r)}\dot{r}^2+1+\frac{L^2}{r^2}-e^{-\nu(r)}E^2=0.
\end{equation}
From the equation of motion
\begin{equation}
 \dot{r}^2  + V(r) = 0,
\end{equation}
we may deduce
\begin{equation}
 V(r) = - e^{-\lambda(r)}\left( e^{-\nu(r)}E^2
  - \frac{L^2}{r^2} -1 \right).
  \end{equation}
However, according to Ref. \cite{NSS01}, since we are dealing with
circular orbits, it is more convenient to use Eq. (\ref{E:eff1})
and the effective potential
\begin{equation}\label{E:Veff}
  V_{eff}=1+\frac{L^2}{r^2}-e^{-\nu(r)}E^2.
\end{equation}
Dealing with circular orbits, the following conditions must be
satisfied: $\dot{r}=0$, $V_r = 0$ and $V_{rr} ~>~0$
\cite{sC83}.  The first condition gives directly
\begin{equation}
  E^2=e^{\nu(r)}\left(1+\frac{L^2}{r^2}\right),
\end{equation}
and from Eq. (\ref{E:Veff}), the second condition yields
\begin{equation}\label{E:second}
   \frac{L^2}{r^2}=\frac{1}{2}r\nu'(r)e^{-\nu(r)}E^2,
\end{equation}
or
\begin{equation}
   E^2=\frac{e^{\nu(r)}}{1-\frac{1}{2}r\nu'(r)}
\end{equation}
and
\begin{equation}
   L^2=\frac{\frac{1}{2}r^3\nu'(r)}{1-\frac{1}{2}r\nu'(r)}.
\end{equation}

Turning next to the tangential velocity $v^{\phi}$, we
have \cite{LL75}
\begin{eqnarray}
   (v^{\phi})^2=r^2e^{-\nu(r)}\left[\left(\frac{d\theta}
   {dt}\right)^2+{sin}^2\theta \left(\frac{d\phi}{dt}
   \right)^2 \right]\nonumber\\
   =r^2 e^{-\nu(r)}\left[\left(\frac{d\theta}{ds}\frac{ds}{dt}
   \right)^2+{{sin}^2\theta} \left(\frac{d\phi}{ds}
   \frac{ds}{dt} \right)^2 \right] \nonumber\\
   =r^2 e^{-\nu(r)} (\dot{\theta}^2+{{sin}^2 \theta} \dot{\phi}^2)\frac{1}{\dot{t}^2}.
\end{eqnarray}

By Eq. (\ref{E:Lsquared}),
\begin{equation}
   (v^{\phi})^2=\frac{L^2\widehat{}}{E^2}\frac{1}{r^2}e^{\nu(r)}
\end{equation}
and by Eq. (\ref{E:second}),
\begin{equation}
   (v^{\phi})^2=\frac{1}{2}r\nu'(r).
\end{equation}
Integrating, we obtain
\[
   e^{\nu}=Br^{l},
\]
where $B$ is an integration constant and $l=2(v^{\phi})^{2}$.

Now from Eq. (\ref{E:Veff}),
\begin{equation}
   V_{eff}(r)_{rr}=\frac{6L^2}{r^4}-E^2e^{-\nu}(\nu')^2
    +E^2e^{-\nu}v''.
\end{equation}
 Substituting for $E^2$, $L^2$, and $\nu$, we obtain
 \[
    V_{eff}(r)_{rr}=\frac{2l}{r^2}>0,
 \]
showing the existence of stable orbits.
\\
\\
\\
\textbf{Appendix 2}
\\
\\
\noindent
The Einstein field equations (in geometrized units $G=c=1$) for
the metric (3) are
\begin{eqnarray}
\frac{b^{\prime }(r)}{ r^{2}}&=&8\pi\rho (r),\\
 2\left(1-\frac{b}{r}\right) \frac{f^\prime}{r} -\frac{b}{r^{3}}&=&8\pi
p_r(r),\\ \left(1-\frac{b}{r}\right)\left[ f^{\prime \prime}+
\frac{f^\prime}{r} +{f^\prime}^2 -          \left\{
\frac{b^{\prime }r-b}{2r(r-b)}\right\}\left(f^{\prime} +
\frac{1}{r}\right)\right]&=&8\pi p_{t}(r).
\end{eqnarray}

Using Eqs. (\ref{E:rho}) and (\ref{E:redshift}), we obtain the
following solutions:
\begin{equation}
b(r)= 8 \pi \rho_s r_s^3 \left[\ln \left( 1+\frac{r}{r_s} \right)
+ \frac{1}{\left( 1+\frac{r}{r_s} \right)} \right],
\end{equation}
\begin{eqnarray}
8 \pi   r_s^2  p_r= \frac{ l}{(\frac{r}{r_s})^2} \left[ 1- \frac{
8 \pi \rho_s r_s^2}{\frac{r}{r_s}} \left\{\ln \left(
1+\frac{r}{r_s} \right) + \frac{1}{\left( 1+\frac{r}{r_s} \right)}
\right\} \right] \nonumber \\-\frac{ 8 \pi \rho_s
r_s^2}{(\frac{r}{r_s})^3} \left[\ln \left( 1+\frac{r}{r_s} \right)
+ \frac{1}{\left( 1+\frac{r}{r_s} \right)} \right],
\end{eqnarray}

\begin{eqnarray}
8 \pi   r_s^2  p_t=  \left[ 1- \frac{ 8 \pi \rho_s
r_s^2}{\frac{r}{r_s}} \left\{\ln \left( 1+\frac{r}{r_s} \right) +
\frac{1}{\left( 1+\frac{r}{r_s} \right)} \right\} \right] \times
\nonumber \\ \left[\frac{l^2}{4(\frac{r}{r_s})^2} - \left\{
\frac{\frac{8 \pi \rho_s r_s^2}{ \left( 1+r/r_s \right)^2} -\frac{
8 \pi \rho_s r_s^2}{(r/r_s)^2} \left[\ln \left( 1+\frac{r}{r_s}
\right) + \frac{1}{ 1+r/r_s }\right] }{2\left[ 1- \frac{ 8 \pi
\rho_s r_s^2}{r/r_s} \left\{\ln \left( 1+\frac{r}{r_s} \right) +
\frac{1}{ 1+r/r_s } \right\} \right]}\right\}  \left\{
\frac{l}{2\frac{r}{r_s}} + \frac{1}{\frac{r}{r_s}}\right\}
\right].
\end{eqnarray}

\vspace{1.0cm}

\section*{Acknowledgement} FR and SR would
like to thank the authority of Inter-University Centre for
Astronomy and Astrophysics (IUCAA), Pune, India for their
hospitality during visits under the Associateship Programme where
a part of the work has been done. FR is also grateful to UGC,
India for financial support under its Research Award Scheme.


\begin{thebibliography}{99}

\bibitem{MT88} M.S. Morris, K.S. Thorne, Am. J. Phys.
   {\bf 56}, 395 (1988)

\bibitem{fL05} F.S.N. Lobo, Phys. Rev. D {\bf 71}, 084011 (2005).

\bibitem{pK09} P.K.F. Kuhfittig, Gen. Rel. Grav. {\bf 41},
1485 (2009)

\bibitem{JKG96} G. Jungman, M. Kamionkowski, K. Griest,
Phys. Rep. {\bf 267}, 195 (1996)

\bibitem{KT90} E.W. Kolb, M.S. Turner, {\it The Early
Universe}, Addison Wesley, Redwood City, California (1990)

\bibitem{sF04} S. Fay, Astron. Astrophys. {\bf 413}, 799 (2004);
T. Matos, F.S. Guzman, D. Nunez, Phys. Rev. D {\bf 62}, 061301
(2000); K.K. Nandi, I. Valitov, N.G. Migranov, Phys. Rev. D {\bf
80}, 047301 (2009);  M. Colpi, S.L. Shapiro, I. Wasserman, Phys.
Rev. Lett. {\bf 57}, 2485 (1986)

\bibitem{NSS00} U. Nukamendi, M. Salgado, D. Sudarsky, Phys.
Rev. Lett. {\bf 84}, 3037 (2000); T. Lee, B. Lee, Phys. Rev. D
{\bf 69}, 127502 (2004); F. Rahaman, R. Mondal, M. Kalam, B.
Raychaudhuri, Mod. Phys. Lett. A {\bf 22}, 971 (2007)

\bibitem{MH04} M.K. Mak, T. Harko,
Phys. Rev. D {\bf 70}, 024010 (2004); F. Rahaman, M. Kalam, A.
DeBenedictis, A.A. Usmani, S. Ray, Mon. Not. R. Astron. Soc. {\bf
389}, 27 (2008); K.K. Nandi, A.I. Filippov, F. Rahaman, S. Ray,
A.A. Usmani, M. Kalam, A. DeBenedictis, Mon. Not. R. Astron. Soc.
{\bf 399}, 2079 (2009)

\bibitem{RKCUR12} F. Rahaman, P.K.F. Kuhfittig, K. Chakraborty,
A.A. Usmani, S. Ray, Gen. Rel. Grav. {\bf 44}, 905 (2012)

\bibitem{BHL08} C.G. B{\"o}hmer, T. Harko, F.S.N.
Lobo, Astropart. Phys. {\bf 29}, 386 (2008)

\bibitem{fR12} F. Rahaman et al., Int. J. Theor.
Phys [in press], arXiv: 1207.2145[gr-qc],
DOI:10.1007/s10773-013-1817-7

\bibitem{jN96} J.F. Navarro et al., Astrophys. J. {\bf 462}, 563 (1996);
J.F. Navarro et al., Astrophys. J. {\bf 490}, 493 (1997)

\bibitem{NSS01} U. Nucamendi, M. Salgado, D. Sudarsky,
  Phys. Rev. D {\bf 63}, 125016 (2001)

\bibitem{Leon1993} J. Ponce de Le\'{o}n, Gen. Relativ. Gravit. {\bf 25}, 1123 (1993)

\bibitem{sC83} S. Chandrasekhar, {\it Mathematical Theory of Black
Holes}, Oxford, Classic Texts (1983)

\bibitem{LL75} L.D. Landau, E.M. Lifshitz, {\it The Classical Theory of Fields}, Oxford, Pergamon Press (1975)

 \end{thebibliography}
\end{document}